\documentclass[preprints,article,accept,moreauthors,pdftex]{Definitions/mdpi}
\usepackage[british]{babel}

\firstpage{1} 
\pubvolume{8}
\issuenum{3}
\articlenumber{145}
\pubyear{2022}
\copyrightyear{2022}
\externaleditor{Academic Editor: Jenni Adams}
\datereceived{19 November 2021}
\dateaccepted{22 February 2022}
\datepublished{24 February 2022}
\hreflink{https://doi.org/10.3390/\linebreak{universe8030145}}

\def\apjl{ApJL}

\def\jcap{JCAP}
\def\mnras{MNRAS}
\def\physrep{Phys. Rep.}
\def\prd{Phys. Rev. D}
\def\prl{Phys. Rev. Lett.}

\newcommand{\centered}[1]{\begin{tabular}{c} #1 \end{tabular}}

\graphicspath{{./plots/}}

\pdfoutput=1

\Title{A 3D Phase Space Analysis of Scalar Field Potentials}

\TitleCitation{A 3D Phase Space Analysis of Scalar Field Potentials}

\Author{Francesco Pace $^{1,2,3,}$*$^{,\dagger}$\orcidA{} and Noemi Frusciante $^{4,\dagger}$\orcidB{}}

\AuthorNames{Francesco Pace and Noemi Frusciante}

\AuthorCitation{Pace, F.; Frusciante, N.}

\address{%
$^{1}$ \quad Dipartimento di Fisica, Universit\`a di Torino, Via P. Giuria 1, I-10125 Torino, Italy\\
$^{2}$ \quad Istituto Nazionale di Fisica Nucleare (INFN), Sezione di Torino, Via P. Giuria 1, I-10125 Torino, Italy \\
$^{3}$ \quad Dipartimento di Fisica ed Astronomia, Alma Mater Studiorum Universit\`a di Bologna, Via Gobetti 93/2, I-40129 Bologna, Italy\\
$^{4}$ \quad Instituto de Astrof\'isica e Ci\^encias do Espa\c{c}o, Faculdade de Ci\^encias da Universidade de Lisboa, Edificio C8, Campo Grande, P-1749016 Lisboa, Portugal; nfrusciante@fc.ul.pt
}

\corres{Correspondence: francesco.pace@unito.it}

\firstnote{These authors contributed equally to this work.}

\abstract{In this study, we present the phase-space analysis of Quintessence models specified by the choice of two potentials, namely the Recliner potential and what we call the broken exponential-law potential, which is a new proposal. Using a dynamical system analysis we provide a systematic study of the cosmological evolution of the two models and their properties. We find new scaling solutions characterised by a constant ratio between the energy density of the scalar field and that of the matter component. These solutions are of high interest in light of the possibility to alleviate the coincidence problem. Additionally, the models also show attractor solutions. We finally construct concrete models built using a double potential according to which one potential realises the early-time scaling regime and the second one allows to exit this regime and to enter in the epoch of cosmic acceleration driven by a scalar-field dominated attractor point.
}


\begin{document}

\section{Introduction}

Since the discovery of the late time cosmic acceleration from Supernovae Ia, a lot of proposals to model this phenomenon have been put forward and tested with the large amount of data we have collected so far. The most successful one \cite{Planck:2018vyg,DES:2017qwj}, based on General Relativity (GR) with the further assumption of an homogeneous and isotropic geometry, is the standard cosmological model (the $\Lambda$CDM model), where the late time acceleration is sourced by the cosmological constant $\Lambda$. The latter is characterised by an equation of state $w_{\Lambda}\equiv P_{\Lambda}/\rho_{\Lambda}=-1$. However, the disagreement between the observed value of $\Lambda$ and its theoretical one, suggested by quantum field theory, has cast doubts on the viability of such scenario. This is known as the cosmological constant problem \cite{Weinberg:1988cp,Martin:2012bt,Joyce:2014kja}. Alternatively, the source of the late-time acceleration can be modelled with a dark energy fluid characterised by a time-dependent equation of state $w_{\rm DE}(t)$. This framework is useful as null test of the $\Lambda$CDM model and enables to identify the tendencies of data with few free parameters.

One easy way to realise a dark energy scenario is to consider a dynamical scalar field $\phi$, and the simplest Lagrangian description is given by Quintessence \cite{Ford1987,Peebles1988,Ratra1988,Caldwell1998,Copeland1998,Steinhardt1999,Martin:2008qp,Tsujikawa:2013fta}. Quintessence is free from theoretical problems such as ghosts and Laplacian instabilities and the late-time acceleration is realised by a slowly varying field along a potential $V(\phi)$. 
This model, besides evolving differently at the background level due to a time-dependent equation of state, also predicts a different growth of perturbations compared to $\Lambda$CDM and its phenomenology depends on the form of the potential. Cosmological data such as galaxy clustering, redshift space distortion, supernovae, baryonic acoustic oscillation, and cosmic microwave background radiation data can be very efficient in constraining the form of the potential \cite{Dutta:2008qn,Chiba:2009gg,Kaiser:1987qv,Wang:1998gt,Linder:2005in}.

In this paper we perform a phase-space analysis \cite{Copeland:2006wr,Zhou:2009cy,Leon:2012mt,Copeland1998,Matarrese:2004xa,Baccigalupi:2000je,Frusciante:2013zop,Albuquerque:2018ymr,Frusciante:2018aew,Frusciante:2018tvu,Barros:2019rdv,Carloni2007,Carloni2010,Carloni2014,Carloni2015,Carloni2016,Bahamonde2018} of the background cosmology of Quintessence. We identify a set of dimensionless variables that allows us to write the background equations as an autonomous system of first-order differential equations. To make the system autonomous, one has to make a choice on the form of the potential. Phase-space analysis in the context of Quintessence has been widely performed, considering many functional forms for the scalar-field potential \cite{Barreiro2000,Jarv2004,Li2005,Ng2001,Ng2001a,UrenaLopez2012,Gong2014,Fang2009,Roy2014,GarciaSalcedo2015,Paliathanasis2015,Matos2009,Zhou2008,Clemson2009}. The simplest option is to consider an exponential potential, as it reduces the dimensionality of the system to be two and its analysis is rather simple \cite{Wiggins2003,Wainwright2005,Amendola2010,Copeland:2006wr}. In this work, we are interested in exploring different forms of the potential. In particular, we will explore the case in which the system is three-dimensional, given the feature of the tracker parameter ($\propto \mathrm{d}^2V/\mathrm{d}\phi^2$) to be a function of the roll parameter ($\propto \mathrm{d}V/\mathrm{d}\phi$). These parameters depend, respectively, on the second and first derivative of the potential with respect to the scalar field. We then apply our method to some specific potentials in order to elucidate it. 

The paper is organised as follows. In Section \ref{Sec:Quintessence} we provide an overview of Quintessence equations on a homogeneous, isotropic, and flat space-time. We also present two different forms for the potential we aim at studying in detail. In Section \ref{Sec:Phasespace} we present the general framework of the phase-space analysis, which will be used to study the evolution of the system associated with the Quintessence model. We then specialise the discussion on two potentials. In Section \ref{Sec:Examples}, we show the evolution of the scalar field density ($\rho_{\phi}$) for the two models characterised by an early-time scaling behaviour ($\rho_{\phi}$ proportional to the matter density). We provide the conclusions in Section \ref{Sec:Conclusion}.

\section{Quintessence Models}\label{Sec:Quintessence}
Quintessence models are described by the following action
\begin{equation}
 S = \int\mathrm{d}x^4\sqrt{-g}[R + \mathcal{L}_{\rm Q}(\phi,g_{\mu\nu}) + \mathcal{L}_{\rm m}(\chi_i,g_{\mu\nu})]\,,
\end{equation}
where $g$ is the determinant of the metric $g_{\mu\nu}$, $R$ is the Ricci scalar, and $\mathcal{L}_{\rm Q}$ is the Quintessence Lagrangian defined as
\begin{equation}\label{eqn:LQ}
 \mathcal{L}_{\rm Q} = -\frac{1}{2}\nabla^{\mu}\phi\nabla_{\mu}\phi - V(\phi)\,,
\end{equation}
where $\phi$ is the scalar field, $V(\phi)$ its potential, and $\nabla_\mu$ is the covariant derivative. Finally, $\mathcal{L}_{\rm m}$ is the Lagrangian of matter fields, $\chi_i$.

If we consider the flat Friedmann-Lema{\^i}tre-Robertson-Walker (FLRW) metric, given~by:
\begin{equation}
 \mathrm{d}s^2 = -\mathrm{d}t^2 + a(t)^2\delta_{ij}\mathrm{d}x^i\mathrm{d}x^j\,,
\end{equation}
where $a(t)$ is the scale factor, $t$ the cosmic time, $x_i$ the spatial coordinates and $\delta_{ij}$ the Kronecker delta, the Friedmann and Klein-Gordon equations read, respectively,
\begin{align}
 & 3M_{\rm pl}^2H^2 = \rho_{\rm m} + \rho_{\phi}\,, \\
 & \ddot{\phi} + 3H\dot{\phi} + V_\phi = 0\,. \label{eqn:KG}
\end{align}

In the equations above, $M_{\rm pl}^2=1/(8\pi G_{\rm N})$ is the reduced Planck mass squared in natural units where $\hbar=c=1$ and with $G_{\rm N}$ being Newton's constant, dots stand for the derivative(s) with respect to cosmic time, $V_\phi\equiv \frac{\mathrm{d}V}{\mathrm{d}\phi}$, $H\equiv \tfrac{\dot{a}}{a}$ is the Hubble parameter, $\rho_{\rm m}$ represents the sum of the matter densities associated to cold dark matter and baryons. Similarly, we can associate an energy density $\rho_{\phi}$ and a pressure $P_{\phi}$ to the scalar field, which read, respectively,
\begin{equation}
 \rho_{\phi} = \frac{1}{2}\dot{\phi}^2 + V(\phi)\,, \quad 
 P_{\phi} = \frac{1}{2}\dot{\phi}^2 - V(\phi)\,.
\end{equation}

From them we can define the equation of the state of the scalar field
\begin{equation}\label{eqn:wphi}
 w_{\phi} = \frac{P_{\phi}}{\rho_{\phi}} = \frac{\dot{\phi}^2 - 2V(\phi)}{\dot{\phi}^2 + 2V(\phi)}\,.
\end{equation}

Note that for a very shallow potential, where $\dot{\phi}\simeq 0$, $w_{\phi}\simeq -1$, it mimics the case of the cosmological constant.

In these models, the potential $V(\phi)$ has to be specified in order to study the dynamics of the scalar field. In this respect, one can follow two different approaches: the first and most common approach specifies the form of the potential and then solves the Klein-Gordon equation to obtain the evolution of the scalar field and finally of $H$ \cite{Peebles1988} (see also~\cite{Avsajanishvili2014} for an explanation of how to construct the initial conditions); the second approach, instead, assumes a functional form for $w_{\phi}$ and then using the Friedmann equations one can reconstruct $V(\phi)$~\cite{Ellis1991,Paliathanasis2015,Battye2016}. The approaches are, obviously, equivalent.

A fundamental theory which, from the first principles, allows us to specify either the functional form of the potential or of the equation-of-state parameter does not exist. All the functional forms proposed for $w_\phi$ and $V(\phi)$ in the literature are, albeit motivated, simply phenomenological. Here we follow the first approach and we consider two forms of the potential. We will then analyse the resulting physical system using the technique of the phase-space dynamics.

In the present analysis we will focus on the potentials defined as follows:
\begin{itemize}
 \item The Recliner potential~\cite{Bag2018}:
 \begin{equation}\label{eqn:recliner}
  V(\phi) = V_0\left[1+\exp{\left(-\frac{\phi}{\sqrt{6\alpha}M_{\rm pl}}\right)}\right] = 2V_0\left[1+\tanh{\left(\frac{\phi}{2\sqrt{6\alpha}M_{\rm pl}}\right)}\right]^{-1}\,,
 \end{equation}
 where $V_0>0$ and $\alpha>0$ are constant. The Recliner model is characterised by tracker solutions \cite{Bag2018} and as such 
 it belongs to a class of models known as $\alpha$-attractors \cite{Kallosh2013a,Kallosh2013b}. They have been proposed to study inflation and have the interesting feature of linking inflation and the present dark-energy dominated era.
 
 \item The broken exponential-law potential. We propose the following potential:
 \begin{equation}\label{eqn:exp}
  V(\phi) = V_0\frac{\chi^p}{(1+A\chi^q)^{2n}}\,,
 \end{equation}
 where $p$, $q$, and $n$ are positive constants and $A$ arbitrary, and $\chi=\exp{(\phi/M_{\rm pl})}$. It represents a simple approximation of $\alpha$-attractor models and is inspired by a general functional form of the potential defining this class of models \cite{LinaresCedeno2019,GarciaGarcia2018}. To the best of our knowledge, this is the first time such a potential is considered. As we will see in Section~\ref{Sec:Examples}, this model gives results very similar to those of the Recliner model. This can somehow be expected as this potential can recover many sub-classes of the $\alpha$-attractor models, including the Recliner. Therefore, the advantage of this potential is to allow a general and systematic investigation, albeit only formally correct when $\phi/M_{\rm pl}\gg 1$, of several models at the same time.
\end{itemize}

\begin{figure}[H]
 \includegraphics[scale=0.6]{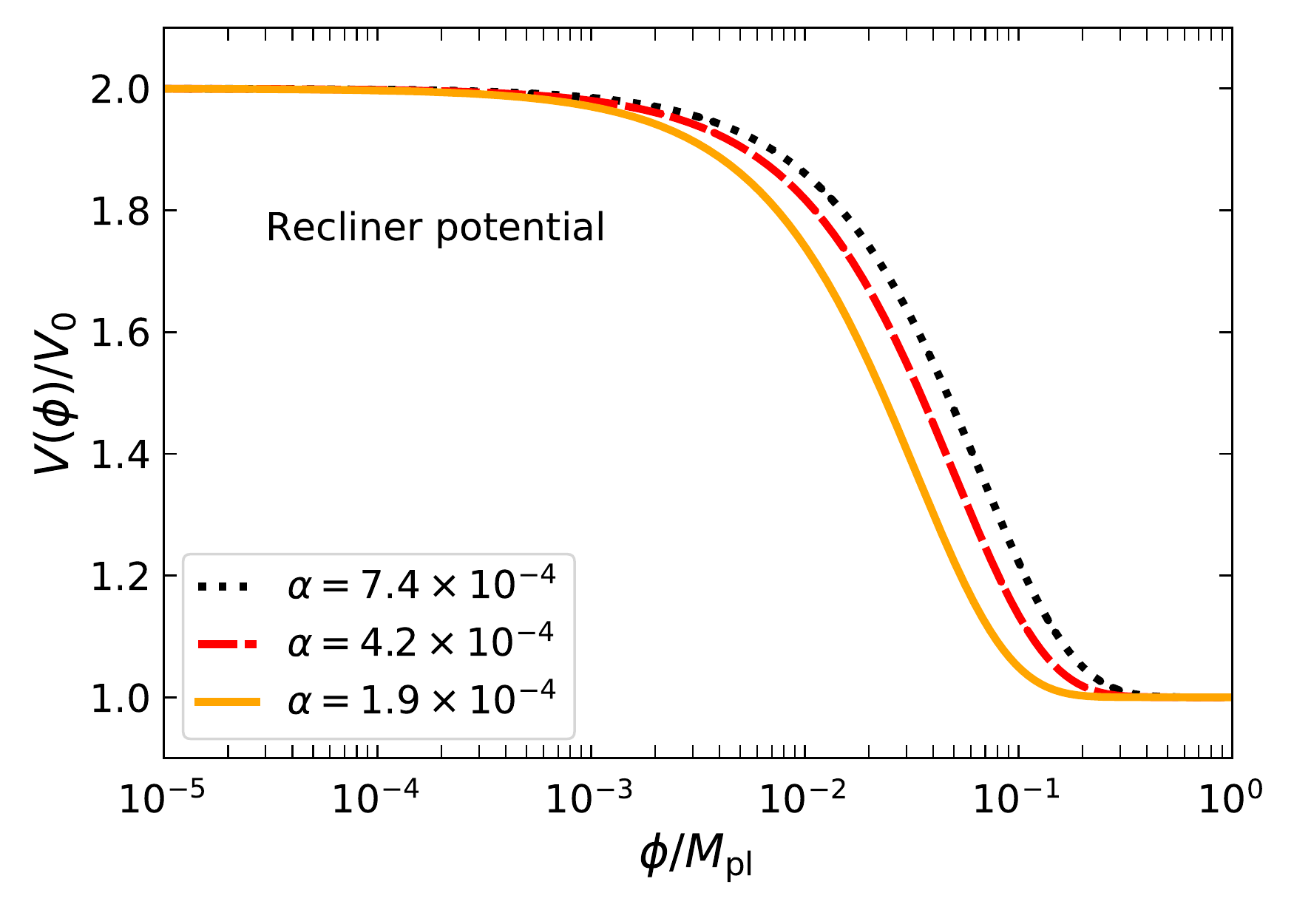}
 \caption{Recliner potential for different values of the parameter $\alpha$ as a function of the scalar field $\phi$. The blue dotted line shows the potential with $\alpha=7.4\times 10^{-4}$, the red dashed line a model with $\alpha=4.2\times 10^{-4}$ and finally the orange solid curve shows a potential with $\alpha=1.9\times 10^{-4}$.}
 \label{fig:Recliner}
\end{figure}

\begin{figure}[H]
 \includegraphics[scale=0.6]{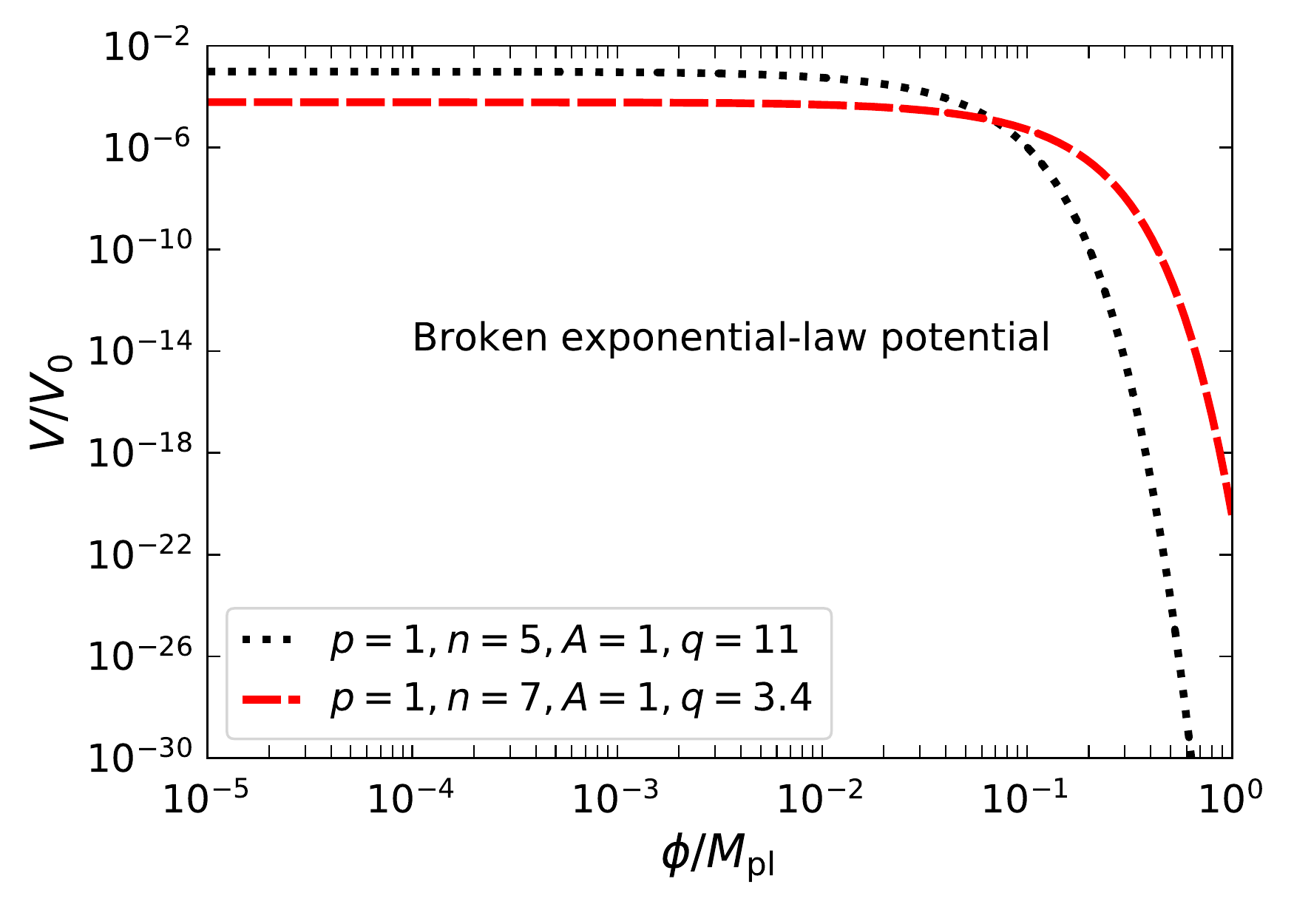}
 \caption{The figure shows the evolution of the broken exponential-law potential for two different combinations of the free parameters as a function of the scalar field $\phi$. The black dotted line shows the potential for $p=1$, $n=5$, $A=1$ and $q=11$; the red dashed line presents a model with $p=1$, $n=7$, $A=1$ and $q=3.4$.}
 \label{fig:BrokenExponential}
\end{figure}

In Figures~\ref{fig:Recliner} and \ref{fig:BrokenExponential} we show the functional form of the Recliner and the broken exponential-law potentials for different sets of the free parameters characterising each model. Different line styles and colours correspond to different parameter sets. For simplicity, we only show positive values of the scalar field $\phi$, but we will also comment on the behaviour of the potentials when $\phi<0$.

\vspace{-6pt}

In Figure~\ref{fig:Recliner} we show the Recliner potential in units of $V_0$. We can individuate three different regimes for positive values of the scalar field: for values $\phi/M_{\rm pl}\ll 1$ (which are characteristic at early times) the potential shows a plateau due to the fact that the exponential function is constant and approximately one. In this limit we have $V(\phi)/V_0\simeq2$. Larger values of $\phi/M_{\rm pl}$ are reached at later times and assuming $\alpha\ll 1$, the contribution of the exponential function is negligible and the Recliner potential flattens again, leading to an equation of state $w_{\phi}\approx -1$, as shown in \cite{Bag2018}. At late times, therefore, $V\approx V_0$. In the intermediate regime, the potential undergoes a transition from $V=2V_0$ to $V=V_0$. For different values of the parameter $\alpha$, the potentials are clearly distinguishable during the tracker regime, whilst behaving in the same way at both early and late times. This shows the attractor nature of this potential: different initial conditions (values of the parameter $\alpha$) will lead to the same final result, regardless of their exact values. For negative values of the scalar field, the evolution of the potential has some peculiar features. For $|\phi/M_{\rm pl}|\ll1$, the plateau at $V\simeq 2V_0$ still exists and for smaller values of the scalar field (higher absolute values) the potential diverges as the argument of the exponential function is positive.

In Figure~\ref{fig:BrokenExponential} we show the evolution of the broken exponential-law potential with respect to the scalar field $\phi$. As we said before, this is inspired by a very general form of the $\alpha$-attractor potentials and we can expect a wide variety of behaviours when choosing different values for the four free parameters. For small, positive values of the scalar field, the potential is flat and later it decreases. This happens because the exponential function is of the order unity when the scalar field is small and does not contribute to the evolution of the potential. When the exponential dominates, the potential diminishes. Note that we do not reach any plateau, so in general we do not expect the late-time dynamics to be similar to the cosmological constant. It is rather straightforward to establish limiting behaviours for this potential. When $\phi/M_{\rm pl}\ll 1$, we obtain $V(\phi)/V_0\simeq (1+A)^{-2n}$, while for $\phi/M_{\rm pl}\gg 1$ we have $V(\phi)/V_0\simeq A^{-2n}\chi^{p-2nq}$, with $\chi$ defined above. This means that when $p-2nq>0$ ($p-2nq<0$) the potential grows (decreases). When we consider negative values of the scalar field, the situation is more interesting. For $|\phi/M_{\rm pl}|\ll1$, our figure would just extend to the left as the potential remains flat. For intermediate values the potential grows, reaches a maximum (its exact position and amplitude depend on the chosen parameters), and then declines to zero as $\chi^p$.

\section{Phase-Space Analysis for Quintessence Models}\label{Sec:Phasespace}
The phase-space analysis is often used in cosmology to study the evolution of a dynamical system around critical points, $p_{{\rm c},i}$. These points satisfy the equilibrium condition $\mathrm{d}p/\mathrm{d}N|_{{\rm c},i}=0$. In this section, we derive the equations necessary to the study of the dynamical system, and as such we consider a system made up of matter (either baryons and cold dark matter or radiation) and the scalar field. 
The first step is to rewrite the equations for the background introduced in Section~\ref{Sec:Quintessence} into an autonomous system of first-order differential equations. To this purpose, we make use of the following dimensionless variables:
\begin{equation}\label{eqn:psmcm}
 x \equiv \frac{\dot{\phi}}{\sqrt{6}HM_{\rm pl}}\,, \quad 
 y \equiv \frac{\sqrt{V}}{\sqrt{3}HM_{\rm pl}}\,, \quad 
 \lambda \equiv -\frac{M_{\rm pl}V_{\phi}}{V}\,, \quad 
 \Gamma \equiv \frac{VV_{\phi\phi}}{V_{\phi}^2}\,, \quad 
 \sigma \equiv \frac{\sqrt{\rho_{\rm m}}}{\sqrt{3}HM_{\rm pl}}\,.
\end{equation}

From a physical point of view, $x$ represents the evolution of the kinetic term, $y$ of the scalar field potential, and $\sigma$ describes the matter density. The quantities $\lambda$ and $\Gamma$ are the roll and tracker parameters, respectively \cite{Zlatev1999}, which determine how fast the scalar field rolls down its potential (in other words, what is the shape of the potential) and whether the tracking behaviour exists.
Using these variables we can derive the expressions for the scalar field density parameter $\Omega_{\phi}$, its equation of state $w_{\phi}$ and the total (also called effective) equation of state of the system, $w_{\rm eff}$, which read, respectively:
\begin{equation}\label{eqn:Qmodel}
 \Omega_{\phi} = x^2+y^2\,, \quad 
 w_{\phi} = \frac{x^2-y^2}{x^2+y^2}\,, \quad 
 w_{\rm eff} = w_{\rm m}+(1-w_{\rm m})x^2-(1+w_{\rm m})y^2\,.
\end{equation}

In the previous set of equations, $0\le w_{\rm m} < 1$ is the equation of state of the matter component. As we are considering only a single matter fluid, $w_{\rm m}$ can be either radiation ($w_{\rm m}=1/3$) or dark matter plus baryons ($w_{\rm m}=0$). However the system can be easily extended to two or more matter components.
Moreover, $0\leq \Omega_{\phi}\leq 1$ and both $x$ and $y$ are finite. While in the two-dimensional analysis they both lie inside a semicircle of radius $r=1$, in three dimensions the situation is more complex, as we will show for each potential separately.
Using Friedmann's equations for Quintessence, we finally write the dynamical system as follows
\begin{equation}\label{eqn:dsQ}
 \begin{split}
  \frac{\mathrm{d}x}{\mathrm{d}N} & = -3x+\frac{\sqrt{6}}{2}\lambda y^2+\frac{3}{2}x[(1+w_{\rm m})(1-y^2)+(1-w_{\rm m}) x^2]\,,\\
  \frac{\mathrm{d}y}{\mathrm{d}N} & = -\frac{\sqrt{6}}{2}\lambda xy +\frac{3}{2}y[(1+w_{\rm m})(1-y^2)+(1-w_{\rm m}) x^2]\,,\\
  \frac{\mathrm{d}\lambda}{\mathrm{d}N} & = -\sqrt{6}\lambda^2(\Gamma-1)x\,,
 \end{split}
\end{equation}
where $N=\ln{a}$. The system (\ref{eqn:dsQ}) is non-linear and non-autonomous. For our purpose to make it autonomous, we can proceed following two options: the first is to assume $\Gamma=1$, and in this case it immediately follows that the form of the potential $V$ is fixed and it is found to be an exponential potential \cite{Copeland1998}. Additionally, $\lambda$ is a constant and this makes the system bi-dimensional. The second assumes $\Gamma-1$ to be a function of $\lambda$. In this case the system becomes three-dimensional and gives rise to an interesting phenomenology associated to different forms of the potential $V$. The latter is the procedure we will consider in this work.

In the following we will show that the choice of the potentials in Section~\ref{Sec:Quintessence} will indeed satisfy the condition $\Gamma(\lambda)$ and we will proceed by identifying the relation between $\Gamma$ and $\lambda$. 
Then, to investigate the dynamics of the system (\ref{eqn:dsQ}), we will determine the evolution around the critical points, $p_{{\rm c},i}$. These points, as said before, are characterised by the equilibrium condition $\mathrm{d}p/\mathrm{d}N|_{{\rm c},i}=0$. Therefore, we will solve the following system of algebraic equations $\tfrac{\mathrm{d}x}{\mathrm{d}N} = \tfrac{\mathrm{d}y}{\mathrm{d}N} = \tfrac{\mathrm{d}\lambda}{\mathrm{d}N} = 0$. The values $p_{{\rm c},i}\in\{x_{\rm c}, y_{\rm c}, \lambda_{\rm c}\}$ satisfying this system will be the critical points we are looking for. Let us notice that we will consider only the points with $y_{\rm c}>0$ as all the functions appearing in its definition are positively defined. However, mathematically the system is symmetric with respect to $y$, and also the points with $y_{\rm c}<0$ will satisfy the equilibrium condition. The final step consists in evaluating the stability of the system around the critical points. To this extent, we compute the eigenvalues $\mu_i$ of the Jacobian matrix $\mathcal{M}$.
Critical points can be classified in hyperbolic and non-hyperbolic points according to whether all eigenvalues of the Jacobian matrix have, respectively, $Re(\mu_i)\neq 0$ and in the second case at least one eigenvalue for which $Re(\mu_i) = 0$.
The dynamical system linearised around hyperbolic critical points is a good approximation of the full nonlinear system and small perturbations do not change qualitatively the phase portrait near the equilibrium point. 
For a 3D system one has three eigenvalues for each critical point and the stability depends on the sign of the eigenvalues: if all the eigenvalues are real and negative, the critical point is stable/attractor; if they are all real and positive the point is unstable; if the eigenvalues are real and at least one is positive while the others negative, the point is a saddle; if at least one eigenvalue is real and there are pairs of complex eigenvalues, the stability depends on the sign of their real part: negative real parts will give a stable point, positive real parts an unstable one, and if one is positive and the others are negative the critical point is a saddle point. These are the points we are more interested in as the analysis around them is robust. However, we will face some cases in which one of the eigenvalues is null. In this case, linear stability is not sufficient to infer the stability of the critical points and one has to resort either to a numerical investigation or to other mathematical methods, such as the Lyapunov method or the centre manifold theory \cite{Carr1982,Wiggins2003}. For an application of the latter method to dynamical systems in cosmology, we refer the reader to \cite{Fang2009,Boehmer2012,Pal2019}.

Finally, note that when discussing the phase-space analysis of Quintessence models, we do not need to take into account stability conditions coming from linear perturbations, as they propagate luminally ($c_{\rm s}^2=1$). In addition, in determining the dynamics of the system, the exact value of $V_0$ is irrelevant, as it simplifies in the calculations.

Let us note that when discussing the critical points, we will see that for $x_{\rm c}=y_{\rm c}=0$, the equilibrium condition is satisfied by an arbitrary value of $\lambda$, which we denote as $\lambda_{\rm a}$ in the following tables.

Before concluding the section, we introduce a definition which will be useful in the analysis to simplify the notation. We define the function $f(\lambda)$ as follows
\begin{equation}\label{eq:flambda}
 f(\lambda) = \sqrt{1-\frac{8(1+w_{\rm m})[\lambda^2-3(1+w_{\rm m})]}{(1-w_{\rm m})\lambda^2}}\,.
\end{equation}

For each critical point, we also discuss its existence, whether it leads to acceleration or not and the corresponding value for the scalar field energy density $\Omega_{\phi}$, its equation of state $w_{\phi}$ and the effective equation of state of the system $w_{\rm eff}$. The condition for the scalar field to produce accelerated expansion is $w_{\rm eff}<-1/3$.

\subsection{The Recliner Potential}

Let us consider the Quintessence model for which the potential is chosen of the form of the Recliner potential in Equation~(\ref{eqn:recliner}). It is easy to show that for this model the $\lambda$ and $\Gamma$ parameters can be written as follows:
\begin{align}
 \lambda = &\, \frac{1}{\sqrt{6\alpha}\left[1+\exp{\left(\frac{\phi}{\sqrt{6\alpha}M_{\rm pl}}\right)}\right]}\,,\\
 \Gamma - 1 = &\, \frac{1}{\sqrt{6\alpha}\lambda} - 1\,.
\end{align}

Therefore, we find that $\Gamma(\lambda)$ and the system (\ref{eqn:dsQ}) is indeed autonomous. The corresponding solutions for the equation $\tfrac{\mathrm{d}\lambda}{\mathrm{d}N}=0$, are 
\begin{equation}
 \lambda_{\rm c,1} = 0\,,\qquad 
 \lambda_{\rm c,2} = \frac{1}{\sqrt{6\alpha}}\,.
\end{equation}

Hence, $0\le\lambda_{\rm a}\le\lambda_{\rm c,2}$. 
We include in Table~\ref{tab:cpmodel1} the list of the critical points of the system~(\ref{eqn:dsQ}) and in Table~\ref{tab:stmodel1} the analysis of the stability. Note that we implicitly assumed that $-\infty<\phi/M_{\rm pl}<+\infty$. This range for the scalar field in fact translates into $\lambda_{\rm c,1}\le\lambda\le\lambda_{\rm c,2}$.

\begin{table}[H]
 \caption{Critical points for the dynamical system with the Recliner model. We label with $\lambda_{\rm a}$ an arbitrary point satisfying the equilibrium condition and together with the coordinates of the critical point, we present its existence, whether it leads to an accelerated expansion phase, the scalar field energy density parameter $\Omega_{\phi}$, the equation-of-state parameter for the scalar field and the total one, $w_{\phi}$ and $w_{\rm eff}$, respectively.}
 
 \begin{adjustwidth}{-\extralength}{0cm}
  \newcolumntype{C}{>{\centering\arraybackslash}X}
  \begin{tabularx}{\fulllength}{Cm{4cm}<{\centering}CCCCC}
   \toprule
   \textbf{Point} & \boldmath{$(x_{\rm c}, y_{\rm c}, \lambda_{\rm c})$} & \textbf{Existence} & \textbf{Acceleration} &  \boldmath{$\Omega_{\phi}$} &  \boldmath{$w_{\phi}$} & \boldmath{$w_{\rm eff}$} \\
   \midrule
   
   $P_1$ & $(0,0,\lambda_{\rm a})$ & Always & No & $0$ & $\nexists$ & $w_{\rm m}$ \\
   \midrule
   $P_2$ & $(0,1,0)$ & Always & Yes & $1$ & $-1$ & $-1$\\
   \midrule
   $P_{3,\pm}$ & $(\pm 1,0,0)$ & Always & No & $1$ & $1$ & $1$ \\
   \midrule
   $P_{4,\pm}$ & $(\pm 1,0,\lambda_{\rm c,2})$ & Always & No & $1$ & $1$ & $1$ \\
   \midrule
   $P_5$ & $(\tfrac{\sqrt{6}}{6}\lambda_{\rm c,2},\sqrt{1-\tfrac{\lambda_{\rm c,2}^2}{6}},\lambda_{\rm c,2})$ & $\lambda_{\rm c,2}^2\le6$ & $\lambda_{\rm c,2}^2<2$ & $1$ & $-1+\tfrac{1}{3}\lambda_{\rm c,2}^2$ & $-1+\tfrac{1}{3}\lambda_{\rm c,2}^2$ \\
   \midrule
   $P_6$ & $(\tfrac{\sqrt{6}}{2}\frac{1+w_{\rm m}}{\lambda_{\rm c,2}},\sqrt{\tfrac{3(1-w_{\rm m}^2)}{2\lambda_{\rm c,2}^2}},\lambda_{\rm c,2})$ & $\lambda_{\rm c,2}^2\ge 3(1+w_{\rm m})$ & No & $\frac{3(1+w_{\rm m})}{\lambda_{\rm c,2}^2}$ & $w_{\rm m}$ & $w_{\rm m}$ \\
   \bottomrule
  \end{tabularx}
 \end{adjustwidth}
 \label{tab:cpmodel1}
\end{table}

From the definition of the variables $x$ and $y$ ($x^2+y^2=\Omega_{\phi}$) we easily find that $0\le x^2+y^2\le 1$ (as we will shortly see, several critical points lie on the boundary) and recalling that $y\ge0$, the dynamical system describes the half-cylinder with boundaries $\{x^2+y^2\le 1$, $y\ge0$, $0\le \lambda\le 1/\sqrt{6\alpha}\}$. The usual phase-space for $x$ and $y$ is, thus, augmented by the boundaries $\lambda=\lambda_{\rm c,1}$ and $\lambda=\lambda_{\rm c,2}$. Although the orbits in these boundaries will not be physical, we can determine, by continuity, the behaviour of the orbits lying inside the half-cylinder. Note that, by construction, therefore, the critical points will lie in the boundaries and trajectories connecting the start and end on these boundaries.

\begin{table}[H]
 \caption{Stability analysis for the dynamical system with the Recliner model. We show the values of the eigenvalues $\mu_i$, with $i\in(1,2,3)$ and the condition(s) for stability.}
 \label{tab:stmodel1}
 \newcolumntype{C}{>{\centering\arraybackslash}X}
 \begin{tabularx}{\textwidth}{Cm{5cm}<{\centering}m{5cm}<{\centering}}
  \toprule
  \centered{\textbf{Point}} & \boldmath{$(\mu_1, \mu_2, \mu_3)$} & \centered{\textbf{Stability}} \\
  \midrule
  \centered{$P_1$} & \centered{$(0,-\tfrac{3}{2}(1-w_{\rm m}),\tfrac{3}{2}(1+w_{\rm m}))$} & \centered{Saddle point} \\
  \midrule
  \centered{$P_2$} & \centered{$(-3,0,-3(1+w_{\rm m}))$} & \centered{Non-hyperbolic stable point} \\
  \midrule
  \centered{$P_{3,+}$} & \centered{$(-\tfrac{1}{\sqrt{\alpha}},3,3(1-w_{\rm m}))$} & \centered{Saddle point} \\
  \midrule
  \centered{$P_{3,-}$} & \centered{$(+\tfrac{1}{\sqrt{\alpha}},3,3(1-w_{\rm m}))$} & \centered{Unstable point} \\
  \midrule
  \centered{$P_{4,+}$} & \centered{$(3(1-\tfrac{1}{6\sqrt{\alpha}}),\tfrac{1}{\sqrt{\alpha}},3(1-w_{\rm m}))$} & \centered{Unstable point for $\alpha\ge\tfrac{1}{36}$\\
  Saddle point for $0<\alpha<\tfrac{1}{36}$} \\
  \midrule
  \centered{$P_{4,-}$} & \centered{$(3(1+\tfrac{1}{6\sqrt{\alpha}}),-\tfrac{1}{\sqrt{\alpha}},3(1-w_{\rm m}))$} & \centered{Saddle point} \\
  \midrule
  \centered{$P_5$} & \centered{$(\lambda_{\rm c,2}^2,\tfrac{\lambda_{\rm c,2}^2}{2}-3,\lambda_{\rm c,2}^2-3(1+w_{\rm m}))$} & \centered{Saddle point for $\alpha\ge\tfrac{1}{36}$} \\
  \midrule
  \centered{$P_6$} & \centered{$(3(1+w_{\rm m}),-\tfrac{3}{4}(1-w_{\rm m})(1\pm f(\lambda_{\rm c,2})))$} & \centered{Saddle point} \\
  \bottomrule
 \end{tabularx}
\end{table}

We now discuss the properties of the critical points in detail:
\begin{itemize}
 \item $P_1$: matter point.\\
 This non-isolated critical point exists for any allowed value of $w_{\rm m}$ and $\lambda$, namely, $0\le w_{\rm m} < 1$ and $0\le\lambda\le\lambda_{\rm c,2}$, and it represents a matter dominated ($\Omega_{\rm m}=1$) solution. More precisely, this is a set of critical points forming a, normally hyperbolic, critical line. Therefore, the stability conditions for this point can be analysed within linear theory. $P_1$ has one null eigenvalue. Indeed, we have:
 \begin{equation}
  \mu_1=0\,,\quad 
  \mu_2=-\frac{3}{2}(1-w_m)\,,\quad 
  \mu_3=\frac{3}{2}(1+w_m)\,,
 \end{equation}
 with $\mu_2<0$ and $\mu_3>0$. The eigenvectors are, respectively:
 \begin{eqnarray*}
  &&\vec{\mathbf{u}}_1=(0,0,1)\,,\\
  &&\vec{\mathbf{u}}_2=(-3\sqrt{3\alpha/2}(1-w_{\rm m})/[\lambda_{\rm a}(6\sqrt{\alpha}\lambda_{\rm a}-\sqrt{6})],0,1)\,,\\ &&\vec{\mathbf{u}}_3=(0,1,0)\,.
 \end{eqnarray*}
 
 From this it follows that on the eigenspace of the zero eigenvalue $\mu_1$, the solutions are time independent, i.e., this eigenspace  is a line of equilibrium. Being the two eigenvalues $\mu_2$ and $\mu_3$ of opposite sign we can conclude that it will be a saddle line of equilibria. This point can represent the early time phase of matter domination.
 \item $P_2$: scalar field point.\\
 This point represents a scalar field dominated solution with $\Omega_{\phi}=1$ and $w_{\phi}=w_{\rm eff}=-1$. 
 This critical point has one null eigenvalue as in the previous case, indeed we have:
 \begin{equation}
  \mu_1=-3\,,\quad \mu_2=0\,,\quad \mu_3=-3(1+w_{\rm m})\,,
 \end{equation}
 with both $\mu_1$ and $\mu_3$ negative. The eigenvectors are, respectively: 
 \begin{eqnarray*}
  &&\vec{\mathbf{u}}_1=(1,0,0)\,,\\
  &&\vec{\mathbf{u}}_2=(\sqrt{6}/6,0,1)\,,\\ 
  &&\vec{\mathbf{u}}_3=(0,1,0)\,.
 \end{eqnarray*}
 This point is, therefore, a non-hyperbolic critical point and linear theory cannot be used to establish its stability. Using the centre manifold theorem, we find that it is a stable point. Details are shown in Appendix \ref{sect:appendix}.
 \item $P_{3,\pm}$: Stiff matter point.\\
 These are two points, labelled $P_{3,\pm} = (\pm 1,0,0)$, having the same background properties but satisfying different stability conditions. They represent scalar field dominated solutions ($\Omega_\phi=1$) with $w_{\rm eff}=1$. 
 In more detail, $P_{3,+}$ is a saddle point and $P_{3,-}$ is an unstable one. The latter corresponds to a stiff matter solution, which is expected to be relevant at early time during inflation. $P_{3,-}$ is the only point of cosmological interest between the two;
 \item $P_{4,\pm}$: Stiff matter point.\\
 The phenomenology of these critical points is very similar to that of points $P_{3\pm}$: $\Omega_\phi=1$ and $w_{\rm eff}=1$. They correspond to: $P_{4,\pm} = (\pm 1,0,\lambda_{\rm c,2})$, with different stability configurations. According to the eigenvalues, we have that $P_{4,-}$ is a saddle point, $P_{4,+}$ can be either unstable for $\alpha\ge1/36$ or a saddle point for $0<\alpha<1/36$. The unstable branch of $P_{4,+}$ is a stiff matter solution and as such it is relevant at early time as for $P_{3,-}$;
 \item $P_{5}$: Scalar field point.\\
 This is another scalar-field-dominated critical point with $x^2+y^2=\Omega_{\phi}=1$. In this case, the equation of state of the scalar field and the total one are $w_{\phi}=w_{\rm eff}=\lambda_{\rm c,2}^2/3-1$, therefore, acceleration is possible for $\lambda_{\rm c,2}^2<2$. However, this point is not an attractor for the system, being a saddle point. It cannot be unstable because it would require $\alpha<1/36$, thus violating the existence condition, i.e. $\alpha\ge 1/36$;
 \item $P_{6}$: Matter scaling point.\\
 This point is characterised by $w_{\rm eff}=w_{\phi}=w_{\rm m}$, which exists for $\lambda_{\rm c,2}^2\ge 3(1+w_{\rm m})$. This point has the density of the scalar field, which scales as the matter one and as such it is said to be a scaling critical point. When $\lambda_{\rm c,2}^2=3(1+w_{\rm m})$, we have $\Omega_{\phi}=1$. These points are of interest to alleviate the Coincidence problem as the initial conditions for the scalar field are fixed and set to follow the matter field. Stability analysis identifies it to be a saddle point. It is of interest to investigate in more detail the stability conditions of $P_6$ for particular choices of $\alpha$ and $w_{\rm m}$. One of these choices is: $w_{\rm m}=0$ and $\alpha=1/36$. This combination of parameters is appropriate, as it stays in the existence domain, i.e, $0<\alpha\le 1/18 $. In this case, $\lambda_{\rm c,2}=\sqrt{6}$ and $P_6$ has coordinates $\{1/2,1/2,\sqrt{6}\}$ and $\Omega_{\phi}=1/2$. The corresponding three eigenvalues now read $\{3,-\tfrac{3}{4}(1\pm\sqrt{3}\,\imath)\}$ and the point is, obviously, a saddle, but it becomes interesting because both $\mu_2$ and $\mu_3$ are complex and the orbits on the plane $\lambda=\lambda_{\rm c,2}$ spiral towards the critical point. For the choice $\lambda_{\rm c,2}^2=3(1+w_{\rm m})$ it follows $f(\lambda_{\rm c,2})=1$ and $P_6$ is still a saddle with eigenvalues $\{3(1+w_{\rm m}), -\tfrac{3}{2}(1-w_{\rm m}), 0\}$.
\end{itemize}

Let us summarise the results of this section. We found one matter point ($P_1$) and two stiff matter points ($P_3$, $P_4$). We also find one matter scaling point ($P_6$). These critical points are of interest in cosmology for early-time evolution of the system; in particular the scaling point, as it is seen as a possibility to alleviate the coincidence problem. 
Indeed, for this case the density of the scalar field is no longer negligibly small compared to the other matter densities at early-time and the model can be compatible with the energy scale associated with particle physics. Moreover, for scaling solutions the initial conditions (ICs) are completely fixed by the model's parameters.
Two other points ($P_2$, $P_5$) are scalar-field dominated points and under the conditions to have accelerated behaviour and stable configurations, only $P_2$ is an attractor solution and as such of cosmological interest at late-time. Viable models undergoing the transitions from $P_1$ to $P_2$ can be constructed. On the other hand, if we want to consider trajectories evolving from the $P_6$ scaling point towards the attractor point $P_2$, one has to employ a second potential. That is because once the trajectories enter the scaling regime, the scalar field can no longer exit because the potential is too steep, therefore, a second potential is necessary to realise the cosmic acceleration. In Section \ref{Sec:Examples} we will show a concrete example with this configuration.

Finally, it is of interest to compare the results of this section with the critical points and their stability in the case of the popular exponential potential (see, e.g., Ref. \cite{Bahamonde2018}). Except for point $P_2$, the other points exist also for the exponential potential. However, the stability of the critical points for the two potentials differs. In fact, while for the exponential potential the critical points $P_5$ and $P_6$ can be stable under certain conditions, this is not the case for the Recliner potential. For example, for point $P_6$, the presence of the eigenvalue $3(1+w_{\rm m})$ forces the point to be a saddle, while its absence for the exponential potential makes it stable.

\subsection{Broken Exponential-Law Potential}

Finally, we discuss the evolution of the dynamical system defined by the Quintessence model with the potential given in Equation~(\ref{eqn:exp}). For this model, in order to close the system of equations in \eqref{eqn:dsQ}, we use:
\begin{align}
 \lambda = &\, -p + 2nq\left[1-\frac{1}{1+A\exp^q{(\phi/M_{\rm pl})}}\right]\,,\\
 \Gamma - 1 = &\, \frac{(p+\lambda)(p-2nq+\lambda)}{2n\lambda^2}\,.
\end{align}

Then, we end up solving the equation $\lambda^2(\Gamma-1)=0$, which has two solutions:
\begin{equation}
 \lambda_{\rm c,1}=2nq-p\,, \qquad \lambda_{\rm c,2}=-p\,.
\end{equation}

Let us notice that because the three parameters $\{n,p,q\}$ are considered to be positive, it follows $\lambda_{\rm c,1}>\lambda_{\rm c,2}$. However, there is no guarantee that $\lambda_{\rm c,1}$ is positive. 
Since the background behaviour of the critical points does not depend on $\lambda$, we will have pairs of points with the same background properties but which satisfy different stability conditions according to the exact value of $\lambda_{\rm c}$.
Additionally, the two-dimensional phase-space of the exponential potential is now augmented by the critical points for $\lambda$ defining the boundaries $\lambda=\lambda_{\rm c,2}$ and $\lambda=\lambda_{\rm c,1}$ and the phase-space of the dynamical system is, therefore, the half-cylinder with boundaries \{$x^2+y^2\le 1$, $y\ge0$, $-p\le\lambda\le 2nq-p$\}, being $2nq$ the height of this half-cylinder. Also in this case, the scalar field can assume both positive and negative values. In fact, $\phi/M_{\rm pl}\to\infty$ corresponds to $\lambda_{\rm c,1}$ and $\phi/M_{\rm pl}\to-\infty$ corresponds to $\lambda_{\rm c,2}$. Therefore, $-\infty<\phi/M_{\rm pl}<\infty$ translates to $\lambda_{\rm c,2}\le\lambda\le\lambda_{\rm c,1}$.

Even though we have the freedom to choose the values of the free parameters, we cannot set $n=0$, otherwise $\Gamma-1$ diverges. This case, therefore, needs a specific treatment. If we set $n=0$, the broken-exponential-law potential and the relevant quantities $\lambda$ and $\Gamma-1$ read
\begin{align}
 V(\phi) = &\, V_0\left[\exp{\left(\frac{\phi}{M_{\rm pl}}\right)}\right]^p\,, \\
 \lambda = &\, -p\,,\\
 \Gamma - 1 = &\, 0\,.
\end{align}

Let us also note that for $\lambda=-p$ constant, we fall back to the case of the exponential potential. The system is, therefore, bidimensional and we do not study it here, as it has been studied in detail in Ref. \cite{Copeland1998}. We recover the same situation if we set $q=0$. Indeed, if we redefine $V_0=V_0/(1+A)^{2n}$, we obtain exactly the same expressions for $n=0$. Hence, in the following analysis we will not consider the cases $n=0$ nor $q=0$.

In principle, the free parameters could assume any positive or even null value. In this work, we will not impose any restriction on the value of these parameters. This implies that a critical point can have different stability regions depending on the combination of these parameters. We present the study of the background properties of the critical points in Table~\ref{tab:cpmodel2} and their stability in Table~\ref{tab:stmodel2}.

\begin{table}[H]
 \caption{Critical points for the dynamical system with the broken exponential-law. $\lambda_{\rm a}$ represents an arbitrary value of the parameter $\lambda$. For each critical point we present its coordinates, the conditions for existence and acceleration and the corresponding values of the scalar field energy density $\Omega_{\phi}$ and equation of state $w_{\phi}$ and the total equation of state of the system $w_{\rm eff}$.}
 \label{tab:cpmodel2}

 \begin{adjustwidth}{-\extralength}{0cm}
  \newcolumntype{C}{>{\centering\arraybackslash}X}
  \begin{tabularx}{\fulllength}{Cm{4cm}<{\centering}CCCCC}
   \toprule
   \textbf{Point} & \boldmath{$(x_{\rm c}, y_{\rm c}, \lambda_{\rm c})$} & \textbf{Existence} & \textbf{Acceleration} & \boldmath{$\Omega_{\phi}$} & \boldmath{$w_{\phi}$} & \boldmath{$w_{\rm eff}$} \\
   \midrule
   $P_1$ & $(0,0,\lambda_{\rm a})$ & Always & No & $0$ & $\nexists$ & $w_{\rm m}$ \\
   \midrule
   $P_2$ & $(0,1,0)$ & Always & Yes & $1$ & $-1$ & $-1$ \\
   \midrule
   $P_{3,\pm}$ & $(\pm 1,0,\lambda_{\rm c,1})$ & Always & No & $1$ & $1$ & $1$ \\
   \midrule
   $P_{4,\pm}$ & $(\pm 1,0,\lambda_{\rm c,2})$ & Always & No & $1$ & $1$ & $1$ \\
   \midrule
   $P_5$ & $(\tfrac{\sqrt{6}}{6}\lambda_{\rm c,1},\sqrt{1-\tfrac{\lambda_{\rm c,1}^2}{6}},\lambda_{\rm c,1})$ & $\lambda_{\rm c,1}^2\le 6$ & $\lambda_{\rm c,1}^2<2$ & $1$ & $-1+\tfrac{1}{3}\lambda_{\rm c,1}^2$ & $-1+\tfrac{1}{3}\lambda_{\rm c,1}^2$ \\
   \midrule
   $P_6$ & $(\tfrac{\sqrt{6}}{6}\lambda_{\rm c,2},\sqrt{1-\tfrac{\lambda_{\rm c,2}^2}{6}},\lambda_{\rm c,2})$ & $\lambda_{\rm c,2}^2\le 6$ & $\lambda_{\rm c,2}^2<2$ & $1$ & $-1+\tfrac{1}{3}\lambda_{\rm c,2}^2$ & $-1+\tfrac{1}{3}\lambda_{\rm c,2}^2$ \\
   \midrule
   $P_7$ & $(\tfrac{\sqrt{6}}{2}\tfrac{1+w_{\rm m}}{\lambda_{\rm c,1}},\sqrt{\tfrac{3(1-w_{\rm m}^2)}{2\lambda_{\rm c,1}^2}},\lambda_{\rm c,1})$ & $\lambda_{\rm c,1}^2\ge 3(1+w_{\rm m})$ & No & $\tfrac{3(1+w_{\rm m})}{\lambda_{\rm c,1}^2}$ & $w_{\rm m}$ & $w_{\rm m}$ \\
   \midrule
   $P_8$ & $(\tfrac{\sqrt{6}}{2}\tfrac{1+w_{\rm m}}{\lambda_{\rm c,2}},\sqrt{\tfrac{3(1-w_{\rm m}^2)}{2\lambda_{\rm c,2}^2}},\lambda_{\rm c,2})$ & $\lambda_{\rm c,2}^2\ge 3(1+w_{\rm m})$ & No & $\tfrac{3(1+w_{\rm m})}{\lambda_{\rm c,2}^2}$ & $w_{\rm m}$ & $w_{\rm m}$ \\
   \bottomrule
  \end{tabularx}
 \end{adjustwidth}
\end{table}

\vspace{-6pt}

\begin{table}[H]
 \caption{Stability analysis for the dynamical system characterised by a broken power-law potential. We present the eigenvalues and the stability condition for each critical point.}
 \label{tab:stmodel2}
 
 \begin{adjustwidth}{-\extralength}{0cm}
  \newcolumntype{C}{>{\centering\arraybackslash}X}
  \begin{tabularx}{\fulllength}{Cm{8cm}<{\centering}m{8cm}<{\centering}}
   \toprule
   \centered{\textbf{Point}} & \boldmath{$(\mu_1, \mu_2, \mu_3)$} & \centered{\textbf{Stability}} \\
   \midrule
   \centered{$P_1$} & \centered{$(0,-\tfrac{3}{2}(1-w_{\rm m}),\tfrac{3}{2}(1+w_{\rm m}))$} & \centered{Saddle point} \\
   \midrule
   \centered{$P_2$} & \centered{$(-3(1+w_{\rm m}),-\tfrac{3}{2}[1\pm\sqrt{1+\tfrac{2p(2nq-p)}{3n}}])$} & \centered{Stable point for $2nq-p<0$ \\
   Saddle point $2nq-p>0$ \\
   Non-hyperbolic saddle point for $p=0$ \\
   Non-hyperbolic stable point for $p=2nq$} \\
   \midrule
   \centered{$P_{3,+}$} & \centered{$(3(1-w_{\rm m}),-\sqrt{6}q,3(1-\tfrac{\sqrt{6}}{6}\lambda_{\rm c,1}))$} & \centered{Saddle point} \\
   \midrule
   \centered{$P_{3,-}$} & $(3(1-w_{\rm m}),\sqrt{6}q,3(1+\tfrac{\sqrt{6}}{6}\lambda_{\rm c,1}))$ & Unstable point for $2nq-p\ge-\sqrt{6}$ \\
   & & Saddle point for $2nq-p<-\sqrt{6}$ \\
   \midrule
   \centered{$P_{4,+}$} & \centered{$(3(1-w_{\rm m}),\sqrt{6}q,3(1-\tfrac{\sqrt{6}}{6}\lambda_{\rm c,2}))$} & \centered{Unstable point} \\
   \midrule
   \centered{$P_{4,-}$} & \centered{$(3(1-w_{\rm m}),-\sqrt{6}q,3(1+\tfrac{\sqrt{6}}{6}\lambda_{\rm c,2}))$} & 
   \centered{Saddle point} \\
   \midrule
   \centered{$P_5$} & \centered{$(-q\lambda_{\rm c,1},\tfrac{1}{2}(\lambda_{\rm c,1}^2-6),\lambda_{\rm c,1}^2-3(1+w_{\rm m}))$} & \centered{Stable point for $0<2nq-p<\sqrt{3(1+w_{\rm m})}$ \\
   Saddle point otherwise \\
   Non-hyperbolic stable point for $2nq-p=0$ \\
   Non-hyperbolic stable point for $2nq-p=\sqrt{3(1+w_{\rm m})}$} \\
   \midrule
   \centered{$P_6$} & \centered{$(q\lambda_{\rm c,2},\tfrac{1}{2}(\lambda_{\rm c,2}^2-6),\lambda_{\rm c,2}^2-3(1+w_{\rm m}))$} & \centered{Stable point for $0<p<\sqrt{3(1+w_{\rm m})}$ \\
   Saddle point for $\sqrt{3(1+w_{\rm m})}<p\le\sqrt{6}$ \\
   Non-hyperbolic saddle point for $p=0$ \\
   Non-hyperbolic stable point for $p=\sqrt{3(1+w_{\rm m})}$} \\
   \midrule
   \centered{$P_7$} & \centered{$(-\tfrac{3q(1+w_{\rm m})}{\lambda_{\rm c,1}},-\tfrac{3}{4}(1-w_{\rm m})[1\pm f(\lambda_{\rm c,1})])$} & \centered{Stable point for $2nq-p>\sqrt{3(1+w_{\rm m})}$ \\
   Saddle point for $2nq-p<-\sqrt{3(1+w_{\rm m})}$ \\
   Non-hyperbolic stable point for $2nq-p=\sqrt{3(1+w_{\rm m})}$} \\
   \midrule
   \centered{$P_8$} & \centered{$(-\tfrac{3q(1+w_{\rm m})}{\lambda_{\rm c,2}},-\tfrac{3}{4}(1-w_{\rm m})[1\pm f(\lambda_{\rm c,2})])$} & \centered{Saddle point} \\
   \bottomrule
  \end{tabularx}
 \end{adjustwidth}
\end{table}

Let us now discuss in more detail each critical point:
\begin{itemize}
 \item $P_1$: Matter point.\\
 The point $P_1$ represents a matter-dominated solution with $\Omega_{\rm m}=1$ and $w_{\rm eff}=w_{\rm m}$, and $\lambda_{\rm a}$ can assume any allowed value, i.e., $\lambda_{\rm c,2}\le\lambda_{\rm a}\le\lambda_{\rm c,1}$. It describes a line of hyperbolic critical points (critical line). The stability requirements identify it as a saddle point. Therefore this point is of interest in the early stage of the universe;
 \item $P_2$: Scalar field point.\\ 
 This point represents a scalar-field dominated solution with $\Omega_{\phi}=1$ and equations of state $w_{\phi}=w_{\rm eff}=-1$ which leads to an accelerated expansion. Depending on the sign of the quantity $2nq-p$ it is a stable or a saddle point. The configuration that is of interest for cosmological purpose is the one for which $2nq-p<0$, which leads to a late-time attractor solution. When $p(2nq-p)=0$, $P_2$ becomes a non-hyperbolic critical point and to investigate its stability we use centre manifold method, as shown in Appendix \ref{sect:appendix}. We find that it is either a stable ($p=2nq$) or an unstable ($p=0$) point;
 \item $P_{3\pm}$ and $P_{4\pm}$: Stiff matter points.\\ 
 These points represent matter-dominated solutions with $\Omega_{\rm m}=1$ and equation of state $w_{\phi}=w_{\rm eff}=1$. This is the typical behaviour of stiff matter. They are never stable points, but they can be either saddle or unstable points, depending on the values of the free parameters as shown in Table~\ref{tab:stmodel2}. In more detail, $P_{3,+}$ and $P_{4,-}$ are always saddle points and $P_{4,+}$ is always unstable; $P_{3,-}$, instead, can be either unstable or a saddle, according to the value of $2nq-p$, see Table~\ref{tab:stmodel2}. The saddle configurations are those of cosmological interest at early time;

 \item $P_5$: Scalar field point.\\ 
 This point exists for $\lambda_{\rm c,1}^2\le6$. It represents another scalar-field dominated solution with $\Omega_{\phi}=1$ and $w_{\phi}=w_{\rm eff}=\lambda_{\rm c,1}^2/3-1$ and when $\lambda_{\rm c,1}=0$, this point coincides with $P_2$. When $\lambda_{\rm c,1}^2<2$ it defines a solution with an accelerated expansion. 
 The stability conditions for $P_5$ are strongly dependant on the values of $\lambda_{\rm c,1}$. It is a stable point if $0\le(2nq-p)\le\sqrt{3(1+w_{\rm m})}$; otherwise it is a saddle point. If $\lambda_{\rm c,1}=2nq-p=0$ or $\lambda_{\rm c,1}=2nq-p=\sqrt{3(1+w_{\rm m})}$, $P_5$ is a stable non-hyperbolic critical point, as shown using the centre manifold theorem. The stable condition is of interest for cosmological purpose as it can define a late-time attractor;
 \item $P_6$: Scalar field point.\\ 
 This point exists for $\lambda_{\rm c,2}^2\le6$. It is another scalar-field dominated point which satisfies the accelerated condition for $p^2<2$. When $\lambda_{\rm c,2}=0$, this point coincides with $P_2$. The point is stable for $0<p<\sqrt{3(1+w_{\rm m})}$. The stable configuration has the requirements to be an attractor solution at late time. When $p=0$ and $p=\sqrt{3(1+w_{\rm m})}$ it is a non-hyperbolic critical point. It is unstable for $p=0$ and stable for $p=\sqrt{3(1+w_{\rm m})}$.
 \item $P_7$ and $P_8$: Matter scaling points.\\
 Both critical points represent matter scaling solutions being the energy density of the scalar field $\Omega_{\phi}=3(1+w_{\rm m})/\lambda_{\rm c}^2$ and $w_{\phi}=w_{\rm eff}=w_{\rm m}$ with $\lambda_{\rm c}$ being either $\lambda_{\rm c,1}=2nq-p$ ($P_7$) or $\lambda_{\rm c,2}=-p$ ($P_8$). The point exists if $\lambda_{\rm c}^2\ge 3(1+w_{\rm m})$. Whilst $P_8$ can only be a saddle point, the phenomenology of $P_7$ is much richer and the point can undergo differ stability configurations. In particular it can be a saddle point for $2nq-p<-\sqrt{3(1+w_{\rm m})}$ and a stable one when $2nq-p>\sqrt{3(1+w_{\rm m})}$. $P_7$ can also be a non-hyperbolic point when $2nq-p=\sqrt{3(1+w_{\rm m})}$. In this case $P_7$ is stable. For both points, it is possible to realise the situation in which $f(\lambda_{\rm c})$ is complex (i.e., when the argument of Equation (\ref{eq:flambda}) is negative), in this case the orbits spiral towards the critical~point.
\end{itemize}

In summary, the model characterised by the broken exponential-law allows for one matter point ($P_1$) and two stiff matter points ($P_3$, $P_4$) that are of interest for the early time evolution of the system. Then we also find two matter scaling points ($P_7$, $P_8$) which can be used to set the evolution of the scalar field to follow that of the matter component at early time. This is of cosmological interest to alleviate the coincidence problem. Three other points ($P_2$, $P_5$, $P_6$) are scalar field dominated solutions and have the proper features to be late-time attractors. While the critical point $P_2$ has the same configuration of the Quintessence model with a Recliner potential, the other two points are not in common. Under the conditions discussed in this section it is possible to realise a viable evolution for the system with trajectories starting from $P_1$ to any of the scalar field dominated critical points. Instead what is difficult to realise is the condition to have trajectories from one of the matter scaling points to a scalar field dominated critical point. This is the same situation present for the Recliner potential. In Section \ref{Sec:Examples} we will show that it is possible to overcome this issue if one considers a second potential. In this case, while the first potential controls the scaling behaviour, the second one allows to exit such a regime and to reach the scalar field dominated critical point.

%-----------------
\section{Working Examples}\label{Sec:Examples}
%----------------

In this section we will show some concrete examples for both potentials we have previously analysed. In particular, we will show how it is possible to realise an evolution for the scalar field which first follows a scaling behaviour with the matter component and then gives rise to the late-time accelerated expansion. For this purpose, we have to consider a double potential, namely $V(\phi)=V_1(\phi)+V_2(\phi)$. That is because for the cases considered in this work it is hard for a single potential to have the scaling behaviour at an early time and later to source the accelerated expansion. In this regard each potential will generate one of the two evolutions. This is a typical procedure already adopted in the presence of exponential potentials \cite{Barreiro2000,Albuquerque:2018ymr}. Therefore, we introduce the following parameters:
\begin{equation}
 y_i = \frac{\sqrt{V_i(\phi)}}{\sqrt{3}HM_{\rm pl}}\,, \quad 
 \lambda_i= -\frac{M_{\rm pl}{V_i}_{\phi}}{V_i}\,, \quad 
 \Gamma_i = \frac{V_i{V_i}_{\phi\phi}}{{V_i}_{\phi}^2}\,,
\end{equation}
for $i=1,2$.

Equation~(\ref{eqn:Qmodel}) then modifies as follow:
\begin{eqnarray}
 && \Omega_{\phi} = x^2 + {y_1}^2 + {y_2}^2\,, \quad 
 w_{\phi} = \frac{x^2-\left({y_1}^2 + {y_2}^2\right)}{x^2+{y_1}^2 + {y_2}^2}\,, \\
 && w_{\rm eff} = w_{\rm m} + (1+w_{\rm m})x^2 - (1+w_{\rm m})\left({y_1}^2 + {y_2}^2\right)\,.
\end{eqnarray}

The same holds for the system (\ref{eqn:dsQ}), which becomes 

\begin{adjustwidth}{-\extralength}{0cm}
%\centering %% If there is a figure in wide page, please release command \centering
\begin{equation}\label{eq:sys2}
\begin{array}{rcl}
\vspace{6pt}
 \frac{\mathrm{d}x}{\mathrm{d}N} &= & \, -3x + \frac{\sqrt{6}}{2}\left(\lambda_1{y_1}^2 + \lambda_2{y_2}^2\right) + \frac{3}{2}x\left[(1+w_{\rm m})\left(1 - {y_1}^2 - {y_2}^2\right) + (1-w_{\rm m})x^2\right] \,,\\\vspace{6pt}
 \frac{\mathrm{d}y_1}{\mathrm{d}N} &= &\, -\frac{\sqrt{6}}{2}\lambda_1 x y_1 + \frac{3}{2}y_1\left[(1+w_{\rm m})\left(1 - {y_1}^2 - {y_2}^2\right) + (1-w_{\rm m})x^2\right] \,, \\\vspace{6pt}
 \frac{\mathrm{d}y_2}{\mathrm{d}N} &= &\, -\frac{\sqrt{6}}{2}\lambda_2 x y_2 + \frac{3}{2}y_2\left[(1+w_{\rm m})\left(1 - {y_1}^2 - {y_2}^2\right) + (1-w_{\rm m})x^2\right] \,, \\\vspace{6pt}
 \frac{\mathrm{d}\lambda_1}{\mathrm{d}N} &= &\, -\sqrt{6}{\lambda_1}^2(\Gamma_1-1)x\,,\\\vspace{6pt}
 \frac{\mathrm{d}\lambda_2}{\mathrm{d}N}& = &\, -\sqrt{6}{\lambda_2}^2(\Gamma_2-1)x\,.
\end{array}
\end{equation}
\end{adjustwidth}

Hereafter, we will consider $w_{\rm m}=0$ and we will neglect the radiation contribution. This assumption does not change the results of this section. We will set ICs for the system \eqref{eq:sys2} in such a way that $\{x_i,y_{1,i},\lambda_{1,i}\}$ are chosen on the scaling critical point, while $\{y_{2,i},\lambda_{2,i}\}$ are selected in order to have $\Omega_{\phi,0}=0.68$. The initial time is set at $a_i=3.3\times 10^{-4}$.

\begin{figure}[H]
 \includegraphics[scale=0.5]{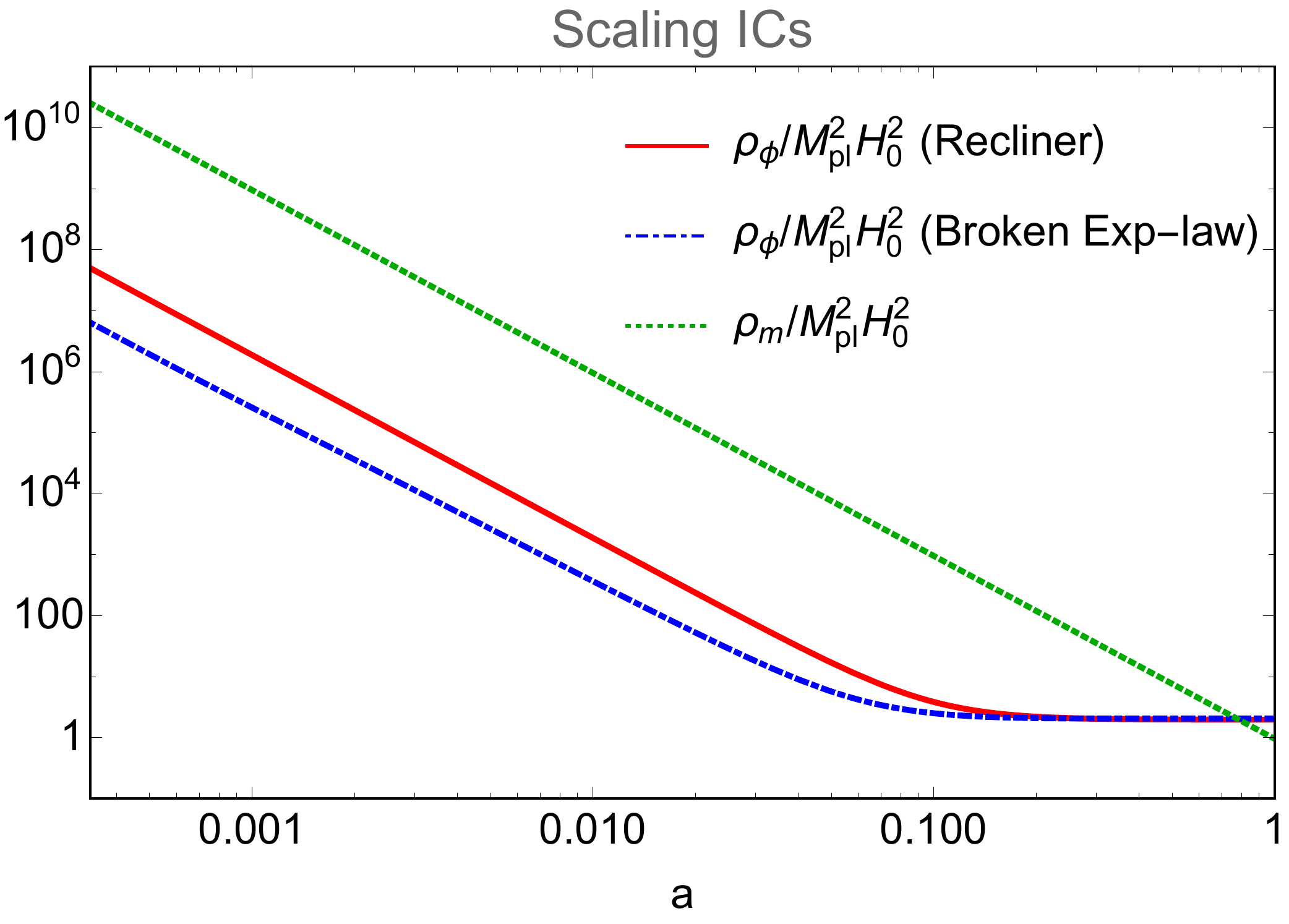}
 \caption{Evolution of the matter density $\rho_{\rm m}$ (green dotted line) and the scalar-field density $\rho_\phi$ for the two types of potentials explored: the Recliner one (red solid line) and Broken exponential-law (blue dot-dashed line). The parameters for the Recliner potential are: $\alpha_1=\times 10^{-4}$, $\alpha_2=1$ and the  ICs are $x_i=y_{1,i}=0.03$, $\lambda_{1,i}=40.83$, $y_{2,i}=0.899\times 10^{-5}$ and $\lambda_{2,i}=0.01$. The parameters for the Broken exponential-law potential are: $p_1=1$, $n_1=5$, $q_1=11$, $p_2=1$, $n_2=7$, $q_2=3.4$ and the ICs are $x_i=y_{1,i}=0.0112362$, $\lambda_{1,i}=109$, $y_{2,i}=0.83\times 10^{-5}$ and $\lambda_{2,i}=-0.7$.}
 \label{fig:Scaling}
\end{figure}

For the Recliner potential case we will then consider a transition between the matter scaling point $P_6$ and the scalar field point $P_2$. The Recliner potential $V_1$ with free parameter $\alpha_1$ will realise the scaling regime and the second Recliner potential $V_2$, with free parameter $\alpha_2$, will then boost the evolution towards the attractor solution. 
In Figure~\ref{fig:Scaling} we show the evolution of the matter density $\rho_{\rm m}$ (green dotted line) and the scalar field density $\rho_\phi$ (red solid line). At early time, the scalar field shows a scaling behaviour $\rho_{\phi} \propto \rho_{\rm m}$. For this case, the field density parameter corresponding to point $P_6$ is $\Omega_\phi (P_6)=18\alpha_1=0.0018$. The latter is consistent with the bound found by the Planck collaboration in 2015 using Cosmic Microwave Background (CMB) radiation measurements, $\Omega_\phi< 0.02$ at 95\% C.L. \cite{Planck:2015bue}. Around $a\sim 0.08$ the scalar field exits the scaling regime and enters in the attractor one. 
In the same Figure \ref{fig:Scaling} we show the evolution for the scalar field density for the Broken exponential-law potential case (blue dot-dashed line). In this case we are considering a transition between the matter scaling point $P_7$ and the scalar field point $P_2$. The Broken exponential-law potential $V_1$ with free parameters $\{p_1,n_1,q_1\}$ will be responsible for the scaling regime and $V_2$ with free parameters $\{p_2,n_2,q_2\}$ for the late-time accelerated expansion. Additionally, in this case the scaling behaviour is evident from the Figure and the field density parameter $\Omega_\phi(P_7)=3/(2n_1q_1-p_1)^2=2.52\times 10^{-4}$. Then, quite early, around $a\sim 2\times 10^{-3}$, it starts to slightly deviate from a strict scaling regime (although very close to it).

Other working examples can be realised by considering the transition $P_1\rightarrow P_2$. In this case we can use for the ICs the values of $x_i,y_{1,i},\lambda_{1,i}$, corresponding to the critical point $P_1$. In the case of the Recliner potential we can use $x_i=y_{1,i}=0$ and an arbitrary value for $\lambda_{1,i}$, which we chose to be $\lambda_{1,i}=1$. The other two ICs are selected such that $\Omega_{\phi,0}=0.68$ as for the previous cases: $y_{i,2}=9\times 10^{-6}$ and $\lambda_{i,2}=10^{-3}$. The latter is the parameter which mostly affects the present day value of $w_{\phi}$, $w_{\phi}^0$. The chosen value allows to have $w_{\phi}^0=-1$, large values of $\lambda_{i,2}$ move it to $w_{\phi}^0>-1$. These ICs allow the trajectories to start on the critical point $P_1$ and to closely approach $P_2$. We found that the values of $\alpha$s do not have any significant impact on the evolution of the matter and scalar field densities for this case. We also ensure that under this condition the matter era is long enough, with a transition to the scalar field dominated era to be at $a= 0.77$. Let us now consider the case of the broken exponential-law potential. The ICs for the transition between $P_1$ and $P_2$ can be chosen as in the previous case, because they share the same coordinates. For this case we fix $p_1=1$, $n_1=5$, $q_1=11$. Then we consider for $P_2$ the configuration for which $p_2=0$ and $q_2=0.1$. For $n_2$ we checked that changing its value of several orders of magnitude ($10^{-7}<n_2<1$) will barely change the general behaviour of the system. However, trajectories closely approach  $P_2$ for the smaller value. For completeness, we select the ICs for $y_{i,2} \approx 0.89\times 10^{-5}$ (to be slightly adjusted depending on $n_2$ in order to have $\Omega_\phi^0=0.68$) and $\lambda_{i,2}=-0.1$. The transition between the matter domination and the scalar field domination happens around $a \approx 0.78$.

We have shown how viable evolution can be obtained with a double potential of the same type. We have tested that scaling regimes realised with any of the two potentials considered in this work can then be followed by any different potential characterised by a late-time stable point, e.g., the exponential potential.

%-----------------
\section{Conclusions}\label{Sec:Conclusion}
%----------------

We have presented a thorough analysis of the phase-space dynamics of two Quintessence models characterised by either the Recliner potential, usually used to model both the late-time accelerated expansion and the inflation one, or by the broken exponential-law potential, which is a novel proposal in this work. We used the formalism of the dynamical systems, which means that the dynamics of the system under consideration needs to be described by an autonomous system of first-order differential equations. In particular, for Quintessence models, the system becomes autonomous if some assumptions on the roll ($\lambda$, proportional to the first derivative of the potential with respect to the scalar field) and tracker ($\Gamma$, proportional to the second derivative of the potential with respect to the scalar field) parameters are made. The standard procedure is to assume $\Gamma=1$ and $\lambda=\mbox{const}$, which corresponds to an exponential potential and the resulting system is bi-dimensional. Otherwise, a different approach considers $\Gamma(\lambda)$ and the system is then three-dimensional. We have considered the latter approach for our analysis, which is more general, and we proceeded in deriving the analytical relations, $\Gamma(\lambda)$, for the two potentials considered.
Let us stress that this approach can be used only if it is possible to find an analytical expression $\phi(\lambda)$, which inserted into $\Gamma$ leads to $\Gamma(\lambda)$ and closes the system. Nevertheless,  this expression might also not be easily managed, and this explains why only a few potentials have been investigated so far.

For each potential we have then identified the critical points and by studying their stability we have determined the evolution of the system around them, and as such the evolution of the Universe governed by a Quintessence field during different cosmological phases. The important critical points for both the systems are the matter scaling solutions ($P_6$ for the Recliner potential and $P_7$, $P_8$ for the broken exponential-law potential) and the scalar-field dominated solutions ($P_2$ for the Recliner potential and $P_2$, $P_5$, $P_6$ for the broken exponential-law potential). The former can be used to set initial conditions at early-time and as such it justifies why the scalar-field density is not negligibly small at early-time and can mitigate the coincidence problem, and the latter provides solutions for the late-time accelerated expansion. In order to have solutions for the system that go from a matter scaling point to a scalar-field dominated point one has to adopt a double potential, for which each potential drives the evolution towards the two epochs. We showed some practical examples in Section \ref{Sec:Examples}, where the double potential is the result of the combination of either two Recliner potentials or two broken exponential-law potentials. For these two concrete examples we also found that the early-time scalar field density is consistent with CMB constraints.

We conclude that both proposed models show very interesting features for realising a viable expansion history and alleviating the coincidence problem. Therefore, it will be of great interest to analyse the phenomenology of perturbations and the effects on cosmological observables and to set constraints on the model parameters using the most recent datasets. We leave this analysis for future work.

\vspace{6pt}

%%%%%%%%%%%%%%%%%%%%%%%%%%%%%%%%%%%%%%%%%%
\authorcontributions{F.P and N.F. have equally contributed. All authors have read and agreed to the published version of the manuscript.}

\funding{F.P. acknowledges partial support from the INFN grant InDark and the Departments of Excellence grant L.232/2016 of the Italian Ministry of Education, University and Research (MIUR) and the Grant No. ASI n.2018-23-HH.0. NF is supported by Funda\c{c}\~{a}o para a Ci\^{e}ncia e a Tecnologia (FCT) through the research grants UIDB/04434/2020, UIDP/04434/2020, PTDC/FIS-OUT/29048/2017, CERN/FIS-PAR/0037/2019 and the personal FCT grant ``CosmoTests -- Cosmological tests of gravity theories beyond General Relativity'' with ref.~number CEECIND/00017/2018. The authors acknowledge support from the FCT project ``BEYLA -- BEYond LAmbda''  with ref. number PTDC/FIS-AST/0054/2021.}

\institutionalreview{Not applicable}

\informedconsent{Not applicable}

\dataavailability{Not applicable}

\acknowledgments{This paper is submitted to the Special Issue  entitled ``Large Scale Structure of the Universe'', led by the authors, and belongs to the section ``Cosmology''. The authors thank an anonymous reviewer, whose comments helped to improve the content of this work.}

\conflictsofinterest{The authors declare no conflict of interest.} 

\appendixtitles{yes}

\appendixstart

\appendix

\section{Centre Manifold Theory}\label{sect:appendix}
In this section we briefly outline the centre manifold method \cite{Carr1982,Wiggins2003}. This is a more sophisticated technique with respect to the linearization procedure described in the main body of the work and it is necessary to discuss the stability of the non-hyperbolic critical points. For an application of the centre manifold theory to dynamical systems in cosmology we refer to \cite{Fang2009,Boehmer2012,Bahamonde2018,Pal2019}.

The goal of the centre manifold theory is to reduce the dimensionality of the dynamical system near the critical point and infer the stability of the reduced system. Then the stability conditions which apply for the reduced system near the critical point will apply to the whole system.

Let us suppose that the dynamical system has dimension $n$ and there are $c$ null and $s$ negative eigenvalues such that $n=s+c$ and that the critical point is located at the origin. If this is not the case, one can always make a translation of the reference frame such that the critical point coincides with the origin of the reference frame. Be $\mathbf{x}\in\mathbb{R}^c$ and $\mathbf{y}\in\mathbb{R}^s$. The dynamical system with null eigenvalues can be written in the following form
\begin{align}
 \mathbf{\dot{x}} = &\, A\mathbf{x} + f(\mathbf{x},\mathbf{y})\,, \label{eqn:cm_x}\\
 \mathbf{\dot{y}} = &\, B\mathbf{y} + g(\mathbf{x},\mathbf{y})\,, \label{eqn:cm_y}
\end{align}
with the dot representing the derivative with respect to the free variable and
\begin{equation}
 f(0,0) = g(0,0) = 0\,, \quad \nabla f(0,0) = \nabla g(0,0) = 0\,,
\end{equation}
where $\nabla f$ and $\nabla g$ represent the matrix of the first derivatives of the vector functions $f$ and $g$, respectively. In Equations~(\ref{eqn:cm_x}) and (\ref{eqn:cm_y}) we have the two vector functions $f\in\mathbb{R}^c$ and $g\in\mathbf{R}^s$ and the matrices $A\in\mathbb{R}^c\times\mathbb{R}^c$ and $B\in\mathbb{R}^s\times\mathbb{R}^s$ having, respectively, zero real part and negative real part eigenvalues.

We define the centre of the manifold of the system in Equations~(\ref{eqn:cm_x}) and (\ref{eqn:cm_y}) as
\begin{equation}
 W^c(0) = \{(\mathbf{x},\mathbf{y})\in\mathbb{R}^c\times\mathbb{R}^s| \mathbf{y}=h(\mathbf{x}),|\mathbf{x}|<\delta, h(0)=0, \nabla h(0)=0\}\,,
\end{equation}
for $\delta$ sufficiently small.

For sufficiently small $\mathbf{u}\in\mathbb{R}^c$, the reduced system reads
\begin{equation}\label{eqn:reduced}
 \mathbf{\dot{u}} = A\mathbf{u} + f(\mathbf{u},h(\mathbf{u}))\,.
\end{equation}

From the definition $\mathbf{y}=h(\mathbf{x})$, we have $\mathbf{\dot{y}}=\nabla h(x)\cdot\mathbf{\dot{x}}$ and using (\ref{eqn:cm_x}) and (\ref{eqn:cm_y}) we arrive to a quasi-linear partial differential equation which $h(\mathbf{x})$ has to satisfy:
\begin{equation}\label{eqn:Nx}
 \mathcal{N}(h(\mathbf{x})) := \nabla h(\mathbf{x})[A\mathbf{x} + f(\mathbf{x},h(\mathbf{x}))] - Bh(\mathbf{x}) - g(\mathbf{x},h(\mathbf{x})) = 0\,,
\end{equation}
and once solved, we insert $h(\mathbf{\mathbf{x}})$ in Equation~(\ref{eqn:reduced}) and we can study its properties. In general, there is no analytical solution to this equation, but its knowledge is not necessary as we only need it around the critical point. Often it suffices to assume $h(\mathbf{x})$ of the form $h_i(x_i)=a_ix_i^2 + b_ix_i^3 + \mathcal{O}(x_i^4)$ and to determine the first non-trivial terms of the Taylor expansion. If they are all null, we need to consider higher order terms.

Operationally, let us consider a system in three variables. If the critical point has coordinates $(x_{\rm c}\,, y_{\rm c}\,, \lambda_{\rm c})$, we define new variables such that $X = x - x_{\rm c}$, $Y = Y - y_{\rm c}$ and $Z = \lambda - \lambda_{\rm c}$, then we rewrite the dynamical system in the new variables and we determine its Jacobian matrix $J$, evaluating it in the centre of the reference frame. We then construct the matrix $\mathcal{T}$ whose columns are the eigenvectors of $J$, evaluate its inverse $\mathcal{T}^{-1}$ and introduce new variables $(u\,, v\,, w)$ defined as $(u\,, v\,, w)^{\rm T} = \mathcal{T}^{-1} (X\,, Y\,, Z)^{\rm T}$, where $\mathrm{T}$ indicates transposition. In these new variables, the system reads
\begin{align}
 \begin{pmatrix}
  \dot{u} \\
  \dot{v} \\
  \dot{w}
 \end{pmatrix}
 = 
 \mathcal{T}^{-1}J\mathcal{T}
 \begin{pmatrix}
  u \\
  v \\
  w
 \end{pmatrix}
 + \text{non-linear terms}\,.
\end{align}

The matrix $D=\mathcal{T}^{-1}J\mathcal{T}$ is now in the desired form and we can determine the reduced system, the vector functions $f$ and $g$, and the unknown function $h$.

In the following we will apply this technique to evaluate the stability of the non-hyperbolic critical points of the Recliner and broken-exponential law potentials.

\subsection{The Recliner Potential}
For the Recliner potential, as evinced from Table~\ref{tab:stmodel1}, the only non-hyperbolic critical point is $P_2=(0,1,0)$, whose eigenvalues are $(0,-3,-3(1+w_{\rm m}))$. Therefore, to infer its (in)stability, we now use the centre manifold theorem.

We first construct a new set of variables, $X=x$, $Y=y-1$ and $Z=\lambda$ and construct the new dynamical system in these new variables. After this, we define a new coordinate set $(u, v, w)$ as

\begin{align}
 \begin{pmatrix}
  u \\
  v \\
  w \\
 \end{pmatrix}
 = 
 \begin{pmatrix}
  0 & 0 & 1 \\
  1 & 0 & -\frac{\sqrt{6}}{6} \\
  0 & 1 & 0 \\
 \end{pmatrix}
 \begin{pmatrix}
  X \\
  Y \\
  Z \\
 \end{pmatrix}
 \,,
\end{align}
and the dynamical system now reads
\begin{align}
 \begin{pmatrix}
  u^{\prime} \\
  v^{\prime} \\
  w^{\prime} \\
 \end{pmatrix}
 = 
 \begin{pmatrix}
  0 & 0 & 0 \\
  0 & -3 & 0 \\
  0 & 0 & -3(1+w_{\rm m}) \\
 \end{pmatrix}
 \begin{pmatrix}
  u \\
  v \\
  w \\
 \end{pmatrix}
 + 
 \begin{pmatrix}
  f(u,v,w) \\
  g_1(u,v,w) \\
  g_2(u,v,w) \\
 \end{pmatrix}
 \,.
\end{align}

To apply the centre manifold theorem, we need the three functions $f$, $g_1$, and $g_2$, which~read
\begin{align}
 f = &\, \sqrt{6}u\left(\frac{\sqrt{6}}{6}u+v\right)\left(u-\frac{1}{\sqrt{6\alpha}}\right)\,,\\
 g_1 = &\, -\frac{\sqrt{6}}{2}u - u\left(\frac{\sqrt{6}}{6}u+v\right)\left(u-\frac{1}{\sqrt{6\alpha}}\right) + \frac{\sqrt{6}}{2}u(1+w)^2 \nonumber\\ 
 & +\frac{3}{2}\left(\frac{\sqrt{6}}{6}u+v\right)\left\{(1+w_{\rm m})\left[1-(1+w)^2\right] + (1-w_{\rm m})\left(\frac{\sqrt{6}}{6}u+v\right)^2\right\} \,,\\
 g_2 = &\, 3(1+w_{\rm m})w - \frac{\sqrt{6}}{2}u\left(\frac{\sqrt{6}}{6}u+v\right)(1+w) \nonumber\\
 & + \frac{3}{2}(1+w)\left\{(1-w_{\rm m})\left(\frac{\sqrt{6}}{6}u+v\right)^2 + (1+w_{\rm m})\left[1-(1+w)^2\right]\right\}\,,
\end{align}
and the function $h$, which we take of the form
\begin{align}
 h = 
 \begin{pmatrix}
  a_2u^2 + a_3u^3 + \mathcal{O}(u^4)\\
  b_2u^2 + b_3u^3 + \mathcal{O}(u^4)\\
 \end{pmatrix}
 \,.
\end{align}

Now using Equation~({\ref{eqn:Nx}}), we obtain

\begin{adjustwidth}{-\extralength}{0cm}
\begin{align}
 \mathcal{N}(u) = 
 \begin{pmatrix}
  \left(3a_2-\frac{1}{6\sqrt{\alpha}}\right)u^2 + \left(3a_3-\sqrt{\frac{3}{2\alpha}}a_2+\frac{3+w_{\rm m}-12(1-w_{\rm m})b_2}{4\sqrt{6}}\right)u^3 + \mathcal{O}(u^4)\\
  \frac{1}{4}(1+w_{\rm m})(1+12b_2)u^2 + \left(\frac{\sqrt{6}}{2}w_{\rm m}a_2-\sqrt{\frac{2}{3\alpha}}+3(1+w_{\rm m})b_3\right)u^3 + \mathcal{O}(u^4)\\ 
 \end{pmatrix}
 = 0\,,
\end{align}
\end{adjustwidth}
which is satisfied for
\begin{equation}
 a_2 = \frac{1}{18\sqrt{\alpha}}\,, \quad 
 a_3 = -\frac{\sqrt{6}(6\alpha-1)}{108\alpha}\,, \quad 
 b_2 = -\frac{1}{12}\,, \quad 
 b_3 = -\frac{\sqrt{6}}{108\sqrt{\alpha}}\,.
\end{equation}

The dynamics of the system restricted to the centre manifold is thus
\begin{equation}
 u^{\prime} = -\frac{1}{\sqrt{6\alpha}}u^2 + \left(1-\frac{1}{18 \alpha}\right)u^3 + \mathcal{O}(u^4)\,,
\end{equation}
and the point is stable, being the coefficient of $u^2$ negative \cite{Boehmer2012}. We can then conclude that the $P_2$ point is an attractor for the system for any value of $\alpha$.

\subsection{The Broken Exponential-Law Potential}

The phenomenology of the broken exponential-law potential is much richer and, under certain combinations of the free parameters, we can have many more non-hyperbolic critical points than the Recliner potential, as we already discussed in the main text. Indeed, we can have up to four of such points. We will examine them independently. Given their length we will not report all the expressions, but exclusively the function $\mathcal{N}$ and the matrix relating the coordinates $(X, Y, Z)$ to $(u, v, w)$ and the final function $f(u)$.

Firstly we consider the critical point $P_2 \equiv(0,1,0)$. When $p(p-2nq)=0$, the eigenvalues of the critical point become $(0, -3, -3(1+w_{\rm m}))$. This is exactly the same configuration of the point $P_2$ for the Recliner model investigated in the previous section and the same variables $(X, Y, Z)$ and $(u, v, w)$ can be constructed. The condition $p(p-2nq)=0$ is satisfied either for $p=0$ or for $2nq-p=0$. For the former case, Equation~(\ref{eqn:Nx}) reads

\begin{adjustwidth}{-\extralength}{0cm}
\begin{align}
 \mathcal{N}(u) = 
 \begin{pmatrix}
  \left(3a_2+\frac{\sqrt{6}}{6}q\right)u^2 + \left[3qa_2+3a_3-\frac{2+n(1+w_{\rm m})(1+12b_2)}{4\sqrt{6}n}\right]u^3 + \mathcal{O}(u^4) \\
  \frac{1}{4}(1+w_{\rm m})(1+12b_2)u^2 + \left[\frac{\sqrt{6}}{2}w_{\rm m}a_2+2qb_2+3(1+w_{\rm m})b_3\right]u^3 + \mathcal{O}(u^4) \\
 \end{pmatrix}
 = 0\,,
\end{align}
\end{adjustwidth}
which is satisfied for
\begin{equation}
 a_2 = -\frac{\sqrt{6}}{18}q\,, \quad 
 a_3 = \frac{\sqrt{6}}{36}\frac{1+2nq^2}{n}\,, \quad 
 b_2 = -\frac{1}{12}\,, \quad 
 b_3 = \frac{1}{18}q\,.
\end{equation}

The dynamics of the system restricted to the centre manifold is
\begin{equation}
 u^{\prime} = qu^2 - \left(1+\frac{2}{3}nq^2\right)\frac{u^3}{2n} + \mathcal{O}(u^4)\,.
\end{equation}

We then conclude that for $p=0$, the point $P_2$ is a saddle as $q>0$ by assumption.

When $p=2nq$, we find
\begin{adjustwidth}{-\extralength}{0cm}
\begin{align}
 \mathcal{N} = 
 \begin{pmatrix}
  \left(3a_2-\frac{\sqrt{6}}{6}q\right)u^2 + \left[3a_3-3qa_2-\frac{2+n(1+w_{\rm m})(1+12b_2)}{4\sqrt{6}n}\right]u^3 + \mathcal{O}(u^4)\\
  \frac{1}{4}(1+w_{\rm m})(1+12b_2)u^2 + \left[\frac{\sqrt{6}}{2}w_{\rm m}a_2-2qb_2+3(1+w_{\rm m})b_3\right]u^3 + \mathcal{O}(u^4)\\
 \end{pmatrix}
 = 0\,,
\end{align}
\end{adjustwidth}
for which
\begin{equation}
 a_2 = \frac{\sqrt{6}}{18}q\,, \quad 
 a_3 = \frac{\sqrt{6}}{36}\frac{1+2nq^2}{n}\,, \quad 
 b_2 = -\frac{1}{12}\,, \quad 
 b_3 = -\frac{1}{18}q\,,
\end{equation}

The dynamical equation is now
\begin{equation}
 u^{\prime} = -qu^2 - \left(1+\frac{2}{3}nq^2\right)\frac{u^3}{2n} + \mathcal{O}(u^4)\,.
\end{equation}

We conclude that $P_2$ for $p=2nq$ is stable.

The second point we investigate is $P_5$. The phenomenology of this point is very rich and it can be a non-hyperbolic point for different values of the free parameters. It is a non-hyperbolic point either for $2nq-p=0$ or $2nq-p=\sqrt{3(1+w_{\rm m})}$. In the first case, it coincides with $P_2$ and it is stable as $q>0$. In fact, the restricted dynamical system reads
\begin{equation}
 u^{\prime} = -qu^2 - \left(\frac{1}{2n}+\frac{q^2}{3}\right)u^3 + \mathcal{O}(u^4)\,.
\end{equation}

In the second case, the eigenvalues are $\{0, -q\sqrt{3(1+w_{\rm m})}, -3(1-w_{\rm m})/2\}$. Since the expression for the vector $\mathcal{N}$ and the relevant coefficients are rather complicated, here we present the reduced dynamical system up to second order, which reads
\begin{equation}
 u^{\prime} = -3\sqrt{2}(1+w_{\rm m})\sqrt{1-w_{\rm m}}u^2 + \mathcal{O}(u^3)\,,
\end{equation}
and the point is always a stable attractor.

The point $P_6$ is non-hyperbolic if $p=0$ or if $p=\sqrt{3(1+w_{\rm m})}$. In the first case, the eigenvalues are the same of $P_2$ and it will have similar stability conditions. In fact,
\begin{equation}
 u^{\prime} = qu^2 - \left(\frac{1}{2n}+\frac{q^2}{3}\right)u^3 + \mathcal{O}(u^4)\,,
\end{equation}
and the point is a saddle.

In the second case, the whole analysis returns the following reduced dynamical system
\begin{equation}
 u^{\prime} = -3\sqrt{2}(1+w_{\rm m})\sqrt{1-w_{\rm m}}u^2 - 6(1+w_{\rm m})(1+3w_{\rm m})u^3 + \mathcal{O}(u^4)\,,
\end{equation}
and the point is stable.

$P_7$ can also be non-hyperbolic for $2nq-p=\sqrt{3(1+w_{\rm m})}$ and we find that the reduced dynamical system is
\begin{equation}
 u^{\prime} = -3(1+w_{\rm m})\sqrt{2(1-w_{\rm m})}u^2 + \mathcal{O}(u^3)\,,
\end{equation}
and the point is stable.

\begin{adjustwidth}{-\extralength}{0cm}
\reftitle{References}

\end{adjustwidth}

\end{document}